\documentclass[aps,prl,showpacs,twocolumn,amsmath,amssymb,superscriptaddress,footinbib]{revtex4}

\usepackage[english]{babel}
\usepackage{latexsym}
\usepackage{graphics}
\usepackage{subfigure}
\usepackage{epsfig}
\usepackage{amsfonts}
\usepackage{amssymb}
\usepackage{amsmath}
\usepackage{bbm}

\begin{document}

\title{Effective Abelian and non-Abelian gauge potentials in cavity QED}


\author{Jonas Larson}
\email{jolarson@kth.se} \affiliation{NORDITA, 106 91 Stockholm,
Sweden}
\author{Sergey Levin} \affiliation{St-Petersburg State University, 198504 St-Petersburg, Russia}

\date{\today}

\begin{abstract}
Cavity QED models are analyzed in terms of field quadrature
operators. We demonstrate that in such representation, the problem
can be formulated in terms of effective gauge potentials. In this
respect, it presents a completely new system in which gauge fields
arise, possessing the advantages of purity, high control of system
parameters as well as preparation and detection methods. Studying
three well known models, it is shown that either Abelian or
non-Abelian gauge potentials can be constructed. The non-Abelian
characteristics are evidenced via numerical simulations utilizing
experimental parameters.
\end{abstract}

\maketitle

{\it Introduction}. -- Gauge fields naturally arise when describing
subatomic interactions. Ranging from classical electromagnetism
\cite{jackson} to quantum Hall systems \cite{gaugeqh} and more
recently cold atoms in optical lattices \cite{gaugecoldat}, gauge
theories have deepen our understanding for fundamental processes in
AMO physics. The simplest and most familiar example is found when
considering a charge particle in an electromagnetic field. In this
case, the gauge theory is Abelian, as the vector components of the
gauge field mutually commute. For non-Abelian gauge fields, on the
other hand, the vector field components are non commuting operators.
In general, time-ordering then becomes important, leading to novel
phenomena. With the experimental progress in especially AMO physics,
non-Abelian gauge structures have drawn great interest in recent
years.

Wilczek and Zee showed that by adiabatically changing a Hamiltonian,
possessing degenerate states, effective non-Abelian gauge potentials
can be achieved \cite{wz}. Indeed, the appearance of gauge
potentials for adiabatic evolution had been demonstrated some years
prior by Mead and Truhlar studying molecular systems
\cite{gaugemol}. Here, adiabaticity is a result from separation of
fast electronic and slow nuclear motions in the molecule, {\it i.e.}
the Born-Oppenheimer approximation. Thus, the gauge potential
derives from intrinsic spatial evolution \cite{mead} and not from
explicit time-dependence as utilized in \cite{wz}. Similar
situations emerge for cold atoms interacting with spatially varying
light fields \cite{gaugelaser}. The center of mass motion of the
atom induces an effective gauge potential, and by considering a
coupled four-level tripod atomic system, Ruseckas {\it et al.}
presented a model which exhibits a non-Abelian gauge structure
\cite{ohberg}. The advantage of this cold atom model is the high
controllability of system parameters as well as efficient
preparation and detection methods.

In this Letter we present a completely different system in which
effective gauge potentials appear, with the same assets as for the
cold atom model, {\it i.e.} purity, coherent control of system
parameters, preparation, and detection. In particular, we consider a
single few-level {\it atom} (could be a true atom or a quantum dot
representing an artificial atom) interacting with one or two
quantized cavity modes. By expressing the field in terms of its
quadrature operators, the structure of the Hamiltonian becomes
similar to the ones encountered in molecular models. In other words,
an adiabatic diagonalization renders effective gauge fields. This is
demonstrated by analyzing three well known models, where two are
endowed with non-Abelian properties.

Since two decades, cavity quantum electrodynamics (QED) with single
or few atoms has delivered some of the most striking experimental
results on pure quantum phenomena \cite{cavityqed}. Among others,
entanglement generation \cite{cavityent}, the quantum measurement
problem and the quantum-classical transition \cite{cavitymeas}, and
verification of the graininess of the quantized electromagnetic
field \cite{rabi}. Cavity QED has attracted even more interest in
recent years due to the realization of coherent coupling of single
quantum dots \cite{cavitydot} or Bose-Einstein condensates
\cite{cavityBEC} to a cavity mode. These experiments pave the way
for the possibility of reaching a super-strong coupling regime of
cavity QED.

The general form of our Hamiltonian reads
\begin{equation}\label{ham1}
H=H_f+H_a+H_I
\end{equation}
where
\begin{equation}
\begin{array}{l}
H_f=\displaystyle{\hbar\sum_k\omega_k\hat{a}_k^\dagger\hat{a}_k},\\
H_a=\displaystyle{\sum_{j=1}^NE_j|j\rangle\langle j|},\\
H_I=\displaystyle{\sum_j\bar{d}^{(j)}\cdot\bar{E}(\mathbf{x})}.
\end{array}
\end{equation}
Here, $\omega_k$ is the $k$'th field mode frequency,
$\hat{a}_k^\dagger$ ($\hat{a}_k$) the creation (annihilation) photon
operator of mode $k$, $E_j$ the energy of atomic level $j$, $N$ the
number of atomic states, $\bar{d}^{(j)}$ the dipole operator for
atomic transition $j$, and $\bar{E}(\mathbf{x})$ is the electric
cavity field. In the dipole approximation we set $\mathbf{x}=0$, and
the field can be written as
\begin{equation}
\bar{E}=\sum_k\bar{\varepsilon}_k\mathcal{E}_ki\left(\hat{a}_k-\hat{a}_k^\dagger\right),
\end{equation}
where $\bar{\varepsilon}_k$ is the polarization vector for mode $k$
and $\mathcal{E}_k$ the corresponding field amplitude. Moreover, the
components of the dipole moment become
$d_\alpha^{(j)}=-e|j\rangle\langle j|\alpha|j+1\rangle\langle
j+1|+h.c.$ with $\alpha=x,y,z$, $e$ the electron charge, and $h.c.$
is the Hermitian conjugate. For our purpose, it is convenient to
express the Hamiltonian in its field quadrature operators
\begin{equation}\label{quad}
\hat{X}_k=\displaystyle{\frac{1}{\sqrt{2}}\left(\hat{a}_k+\hat{a}^\dagger_k\right)},\hspace{1cm}
\hat{P}_k=\displaystyle{\frac{i}{\sqrt{2}}\left(\hat{a}_k-\hat{a}^\dagger_k\right)},
\end{equation}
obeying the canonical commutation relations;
$[\hat{X}_k,\hat{P}_{k'}]=i\delta_{kk'}$. The field quadrature
operators (\ref{quad}) are easily measured experimentally. Indeed,
the ENS group of S. Haroche recently presented experimental results
were the full phase space distribution of a cavity mode was assessed
\cite{wigner}. In terms of (\ref{quad}), the Hamiltonian
(\ref{ham1}) takes the form
\begin{equation}\label{ham2}
\begin{array}{lll}
H & = &
\displaystyle{\hbar\sum_k\omega_k\left(\frac{\hat{P}_k^2}{2}+\frac{\hat{X}_k^2}{2}\right)+\sum_{j=1}^NE_j|j\rangle\langle
j|}\\ & &
-\displaystyle{\sum_{j=1}^N\sum_k\bar{d}^{(j)}\cdot\bar{\varepsilon}_kg_{jk}\hat{P}_k},
\end{array}
\end{equation}
where $g_{jk}$ is the effective atom-field coupling between the
transition $j$ and the mode $k$. The general form of the Hamiltonian
(\ref{ham2}) works as a starting point to analyze more specific
cavity QED models.

{\it Rabi model}. -- The simplest non-trivial situation considers a
single cavity mode interacting with one atomic transition.
Experimentally, such idealized situations are accessible by
utilizing high-$Q$ cavities with well resolved resonance frequencies
$\omega_k$, see for example
Refs.~\cite{cavityqed,cavityent,cavitymeas,rabi}. For this model we
leave out the indices $j$ and $k$. Introducing the Pauli matrices
$\hat{\sigma}_x=|2\rangle\langle1|+|1\rangle\langle2|$,
$\hat{\sigma}_y=-i|2\rangle\langle1|+i|1\rangle\langle2|$, and
$\hat{\sigma}_z=|2\rangle\langle2|-|1\rangle\langle1|$, and without
loss of generality assuming that $\bar{d}\cdot\bar{\varepsilon}$ is
purely real, we obtain the Rabi Hamiltonian
\begin{equation}
H_{Rabi}=\hbar\omega\left(\frac{\hat{P}^2}{2}+\frac{\hat{X}^2}{2}\right)+\frac{\hbar\Omega}{2}\hat{\sigma}_z-\hbar g\hat{\sigma}_x\hat{P}.
\end{equation}
Here, $g$ is the vacuum Rabi frequency, giving the effective
atom-field coupling, and we have chosen our zero energy such that
$E_1=-\hbar\Omega/2$ and $E_2=\hbar\Omega/2$.

We note that the Hamiltonian can be rewritten as
\begin{equation}
H_{Rabi}=\hbar\omega\left(\frac{(\hat{P}-\hat{A})^2}{2}+\frac{\hat{X}^2}{2}\right)+\frac{\hbar\Omega}{2}\hat{\sigma}_z+\hat{\Phi},
\end{equation}
where we have introduced the scaled gauge potential
$\hat{A}=\frac{g}{\omega}\hat{\sigma}_x$ and the scalar potential
$\hat{\Phi}=-\hbar\frac{g^2}{2\omega}$. Under a unitary
transformation of the state vector $\Psi(\hat{X},t)$,
\begin{equation}
\Psi(\hat{X},t)\rightarrow U^\dagger(\hat{X},t)\Psi(\hat{X},t),
\end{equation}
the gauge and scalar potential transforms accordingly
\begin{equation}\label{gaugetrans}
\begin{array}{l}
\hat{A}\rightarrow
\displaystyle{U^\dagger(\hat{X},t)\hat{A}U(\hat{X},t)-U^\dagger(\hat{X},t)\frac{\partial}{\partial
\hat{X}}U(\hat{X},t)},\\ \hat{\Phi}\rightarrow
\displaystyle{U^\dagger(\hat{X},t)\hat{\Phi} U(\hat{X},t)-i\hbar
U^\dagger(\hat{X},t)\frac{\partial}{\partial t}U(\hat{X},t)}.
\end{array}
\end{equation}
The above equation demonstrates the gauge structure of the effective
potentials.

It should be observed that the appearance of the effective gauge
potentials results from the non-stationary dynamics of the quantized
cavity field, and not from adiabatic particle motion as in earlier
works on cold atoms and molecular physics
\cite{gaugemol,mead,gaugelaser,ohberg}. Furthermore, note that by
applying the rotating wave approximation, $H_{Rabi}$ turns into the
solvable Jaynes-Cummings Hamiltonian \cite{jonas1} that has served
as workhorse in the field of quantum optics ever since it was
introduced \cite{jc}. Imposing such an approximation would make the
present analysis less transparent since the atom field coupling
would then depend on both $\hat{P}$ and $\hat{X}$ rendering
complicated gauge potentials. In addition, for current quantum dot
cavity QED systems, application of the rotating wave approximation
is not always justified.

\begin{figure}[h]
\centerline{\includegraphics[width=4.7cm]{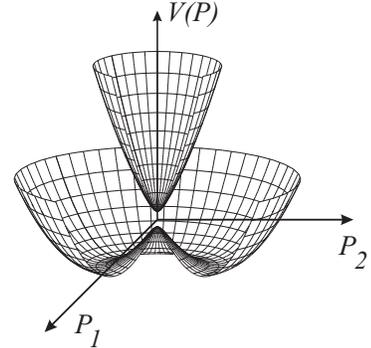}}
\caption{Effective APSs of the bimodal Rabi Hamiltonian
(\ref{hamBR}) in the $g>\sqrt{\omega\Omega}$ case, where the lower
APS attains the sombrero shape. The conical intersection is located
at the origin, $P_1=P_2=0$.} \label{fig1}
\end{figure}

{\it Bimodal Rabi model}. -- In order to acquire non-Abelian gauge
potentials, additional degrees of freedom must be included. This is
easily accomplished in our cavity QED setting by considering a
bimodal cavity field \cite{cavitybimode}. Hence, the atom interacts
simultaneously with two cavity modes. It is worth noting that
experiments on bimodal cavities have been successfully demonstrated
\cite{bimodeexp}. The simplest extension of the Rabi model is to
keep the two-level structure of the atom and simply add one
additional cavity mode. Recently, it was shown that the
corresponding cavity QED model has a Jahn-Teller structure
\cite{jonas2}. For appropriate choices of polarizations and atomic
dipole moments, one obtains the $E\times\varepsilon$ Jahn-Teller
Hamiltonian which has been thoroughly studied in especially
molecular physics \cite{conical}. The model possesses a conical
intersection in which the adiabatic potential surfaces become
degenerate. Encircling the intersection brings about a non-zero
geometrical phase characterized by a gauge potential. We note that
geometrical phases have been discussed in terms of cavity QED
\cite{berry,jonas2}. These, however, do either consider the Abelian
situation or Berry phases originating from a time-dependent Hamiltonian and consequently the gauge field derives from external driving rather than from intrinsic dynamical evolution.

The $E\times\varepsilon$ cavity QED Hamiltonian is given by
\cite{jonas2}
\begin{equation}\label{hamBR}
\begin{array}{lll}
H_{BR} & = &
\displaystyle{\hbar\omega\sum_{k=1,2}\left(\frac{\hat{P}_k^2}{2}+\frac{\hat{X}_k^2}{2}\right)+\frac{\hbar\Omega}{2}\hat{\sigma}_z}\\
& & -\displaystyle{\hbar
g\left(\hat{\sigma}_x\hat{P_1}+\hat{\sigma}_y\hat{P}_2\right)}.
\end{array}
\end{equation}
Here, we have assumed equal mode frequencies and equal atom-field
strengths. Contrary to Jahn-Teller models encountered in molecular
physics, the conical intersection appears in momentum space rather
than in position. In the condensed matter community, a coupling as
the one in Eq.~(\ref{hamBR}) is usually said to be on Rashba form
\cite{rashba}. The adiabatic potential surfaces (APS), defined as
$V_{ad}^\pm(P_1,P_2)=\hbar\omega\left(P_1^2+P_2^2\right)/2\pm\hbar\sqrt{\Omega^2/4+g^2\left(P_1^2+P_2^2\right)}$,
are envisaged in Fig.~\ref{fig1}. Due to the non-zero $\Omega$, the
conical section becomes avoided. The lower APS has a sombrero shape
whenever $g>\sqrt{\Omega\omega}$, otherwise a global minimum at the
origin is exhibited. It follows directly from the form of the
Hamiltonian (\ref{hamBR}), that the vector and scalar potentials
read
\begin{equation}
(\hat{A}_1,\hat{A}_2)=\frac{g}{\omega}(\hat{\sigma}_x,\hat{\sigma}_y),\hspace{1cm}\hat{\Phi}=-\hbar\frac{g^2}{\omega},
\end{equation}
and since $[\hat{A}_1,\hat{A}_2]\neq0$, the gauge potential is
non-Abelian.

\begin{figure}[h]
\centerline{\includegraphics[width=7.5cm]{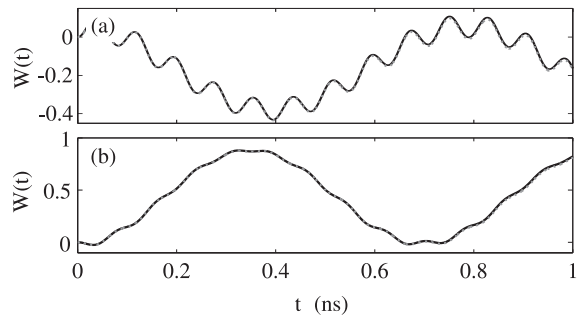}} \caption{Time
evolution of the atomic inversion for clockwise (a) and
anti-clockwise (b) propagation. Solid lines display the ideal case
of no losses, while for the dotted lines both cavity and atomic
losses are included. The non-Abelian structure of the model is
clearly visible. The system parameters are $P_{10}=2$, $P_{20}=0$,
$X_{10}=0$, $X_{20}=\pm5$, $\Omega/2\pi=6.9$ GHz, $\omega/2\pi=5.7$
GHz, $g/2\pi=105$ MHz, $\gamma/2\pi=1.9$ MHz, and $\kappa/2\pi=250$
kHz.} \label{fig2}
\end{figure}

In a recent work \cite{jonas3}, considering a spinor Bose-Einstein
condensate, a proposal for detecting non-Abelian characteristics was
put forwarded. The idea is to initialize a state, such that it is
given by a fairly localized wave packet with non-zero average
momentum, and then boost the wave packet either clockwise or
anti-clockwise. Due to the non-Abelian structure, the two paths
will render different dynamics despite the polar symmetry of the
problem. To demonstrate this, we chose the atom to be initially in a
superposition state $\left(|2\rangle-|1\rangle\right)/\sqrt{2}$ and
the two fields to be in coherent states
\begin{equation}
\psi_i(P_i,0)=\frac{1}{\sqrt{\pi}}e^{-(\Im
P_{i0})^2}e^{-iX_{i0}P_i}e^{-(P_i-P_{i0})^2/2},\hspace{0.3cm}i=1,2,
\end{equation}
where $P_{i0}$ and $X_{i0}$ are initial average momentum and
position respectively. These are related to the initial coherent
state amplitude $\alpha_{i0}=(X_{i0}+iP_{i0})/\sqrt{2}$. For the
situation at hand, we set $P_{10}>0$, $P_{20}=0$, $X_{10}=0$, and
$X_{20}\neq0$. In momentum representation, this gives an initial
wave packet mainly at the lower APS located at the positive
$P_1$-axis and with an initial velocity perpendicular to this axis.
For $X_{20}>0$, the wave packet sets off clockwise around the
origin, while $X_{20}<0$ results in anti-clockwise evolution. Our
numerical simulation utilizes the split-operator method, which gives
the time-evolved wave packet at any instant of time. Solid lines of
Fig.~\ref{fig2} presents the results for the atomic inversion,
$W(t)=p_2(t)-p_1(t)$ where $p_{i}(t)$ is the probability of finding
the atom in the state $|i\rangle$ at time $t$, for clockwise
propagation (a) and for anti-clockwise evolution (b). The final time
corresponds to approximately six roundtrips around the conical
intersection. We point out that the atomic inversion can be
experimentally measured up to a few percent accuracy
\cite{cavityqed}. The discrepancy between the lines of (a) and (b),
evident even at very short time scales ($\sim$ns), is a
manifestation of the underlying non-Abelian character.

A deeper understanding is obtained by studying the evolution of the
fields in phase space. As for a classical harmonic oscillator, both
fields will encircle the origin of phase space with, for the example
of Fig.~\ref{fig2}, radii approximately 2 and 5 respectively. The
effective magnetic field deriving from the gauge potential renders a
momentum dependent force. The momentum is in general different for
clockwise and anti-clockwise evolution, and hence this effective
Lorentz force acting on the field distributions implies slightly
deviating trajectories in phase space for the two cases. This
difference is the origin for the momentum-dependence of the atomic
inversion seen in Fig.~\ref{fig2}.

In realistic situations, both cavity and atomic losses come into
play. We schematically take these losses into account by consider
time evolution of the effective non-hermition Hamiltonian
$H_{eff}=H_{BR}-i\kappa\left(\hat{a}_1^\dagger\hat{a}_1+\hat{a}_2^\dagger\hat{a}_2\right)-i\gamma|2\rangle\langle
2|$. Here $\kappa$ and $\gamma$ are the photon decay rate and the
atomic spontaneous emission rate respectively. The results of such
numerical simulation are given by the dotted lines of
Fig.~\ref{fig2}. The  parameters used in the examples of
Fig.~\ref{fig2} have been chosen according to the quantum dot cavity
QED experiments of Ref.~\cite{cavitydot}. Due to the short
interaction times of Fig.~\ref{fig2}, losses play a minor role. The
importance of losses becomes apparent only for longer time scales.

{\it Bimodal $\Lambda$ model}. -- Our final example considers a
$\Lambda$-atom with two lower meta stable states $|1\rangle$ and
$|2\rangle$ coupled to an excited state $|3\rangle$ via two cavity
fields 1 and 2 respectively. Numerous theoretical works on this
system have been put forward \cite{cavitybimode}, and
Ref.~\cite{lambda} presents some experimental studies of $\Lambda$
cavity experiments.

The Hamiltonian for this system reads
\begin{equation}
\begin{array}{lll}
H_{\Lambda} & = &
\displaystyle{\hbar\omega\sum_{k=1,2}\left(\frac{\hat{P}_k^2}{2}+\frac{\hat{X}_k^2}{2}\right)+\sum_{j=1,2,3}E_j|j\rangle\langle
j|}\\ & & -\displaystyle{\hbar
g\left(\hat{P}_1|3\rangle\langle1|+\hat{P}_2|3\rangle\langle2|+h.c.\right)},
\end{array}
\end{equation}
where we have assumed that the effective atom-field couplings are
real and the same between the two transitions. For degenerate lower
atomic states $E_1=E_2$, the three APSs possess a Renner-Teller
intersection \cite{rt} between the lower and the middle surface and
another one between the middle and the upper surface. These
intersections are characterized by that the tangents of the two
surfaces are the same at the degenerate point. Moreover, the Berry
phase acquired by encircling a Renner-Teller intersection in
position space vanishes. Here, however, the intersection is in the
momentum space giving rise to a non-Abelian Berry phase in position
space, where we have checked that for $E_1=E_2$ the diagonal terms
of the corresponding $3\times3$ geometric phase matrix are zero but
some of the off-diagonal terms are indeed non-zero. If the two lower
atomic states are not degenerate $E_1\neq E_2$, both Renner-Teller
intersections split into two non-avoided conical intersection.

As in the previous bimodal example, the gauge potential is
non-Abelian;
\begin{equation}
(\hat{A}_1,\hat{A}_2)=\frac{g}{\omega}(\hat{\lambda}_4,\hat{\lambda}_6),\hspace{1cm}\hat{\Phi}=\hbar\frac{g^2}{3\omega}\left(1-\frac{\sqrt{3}}{2}\hat{\lambda}_8\right),
\end{equation}
with $\hat{\lambda}_j$ being the Gell-Mann matrices
($\hat{\lambda}_4=|3\rangle\langle1|+|1\rangle\langle3|$,
$\hat{\lambda}_6=|3\rangle\langle2|+|2\rangle\langle3|$, and
$\hat{\lambda}_8=(|1\rangle\langle1|+|2\rangle\langle2|-2|3\rangle\langle3|)/\sqrt{3}$).

{\it Concluding remarks}. -- We have demonstrated how cavity QED
models provide novel systems exhibiting artificial gauge potentials.
This derives from the quantized motion of the cavity fields; the
dynamics of the fields in phase space. Moreover, using numerical
simulations we showed that non-Abelian characteristics should be
detectable under realistic conditions. From this rather unusual
approach to cavity QED, presented in this Letter and earlier in
\cite{jonas1,jonas2}, it is clear that these models are very rich.
We are at the moment analyzing the prospects of achieving
zitterbewegung or Hall effects with cavity QED setups \cite{zitt}.
We hope that the current contribution will encourage experiments
along these lines.

We wish to thank Prof. Erik Sj\"oqvist for fruitful discussions, and
JL acknowledge support from the MEC program (FIS2005-04627).

\end{document}